\documentstyle[12pt]{article}
\newcommand{\bea}{\begin{eqnarray}}
\newcommand{\eea}{\end{eqnarray}}
\newcommand{\be}{\begin{equation}}
\newcommand{\ee}{\end{equation}}
\newcommand{\vs}[1]{\vspace{#1 mm}}

\newcommand{\dsl}{\pa \kern-0.5em /}

\newcommand{\pa}{\partial}

\newcommand{\nn}{\nonumber\\}

\begin{document}
\topmargin 0pt
\oddsidemargin 0mm

\begin{flushright}
USTC-ICTS-04-7\\
MCTP-04-17\\
hep-th/0403147\\
\end{flushright}

\vs{2}
\begin{center}
{\Large \bf
Supergravity approach to tachyon condensation on the brane-antibrane system}
\vs{10}

{\large J. X. Lu$^a$\footnote{E-mail: jxlu@ustc.edu.cn}
 and Shibaji Roy$^b$\footnote{E-mail: roy@theory.saha.ernet.in}}

 \vspace{5mm}

{\em
 $^a$ Interdisciplinary Center for Theoretical Study\\
 University of Science and Technology of China, Hefei, Anhui 230026,
P. R. China\\
and\\
Interdisciplinary Center of Theoretical Studies\\
Chinese Academy of Sciences, Beijing 100080, China\\
 and\\

Michigan Center for Theoretical Physics\\
Randall Laboratory, Department of Physics\\
University of Michigan, Ann Arbor, MI 48109-1120, USA\\

\vs{4}

 $^b$ Saha Institute of Nuclear Physics,
 1/AF Bidhannagar, Calcutta-700 064, India}
\end{center}

\vs{5}
\centerline{{\bf{Abstract}}}
\vs{5}
\begin{small}
We study the tachyon condensation on the D-brane--antiD-brane
system from the supergravity point of view. The non-supersymmetric
supergravity solutions with symmetry ISO($p,1$) $\times$ SO($9-p$)
are known to be characterized by three parameters. By interpreting
this solution as coincident $N$ D$p$-branes and ${\bar N}$ ${\bar
{\rm D}}p$-branes we give, for the first time, an explicit
representation of the three parameters of supergravity solutions
in terms of $N, \bar N$ and the tachyon vev. We demonstrate that
the solution and the corresponding ADM mass capture all the
required properties and give a correct description of the tachyon
condensation advocated by Sen on the D-brane--antiD-brane system.
\end{small}
\newpage

It is well-known that a coincident D-brane--antiD-brane pair (or a
non-BPS D-brane) in Type II string theories is unstable due to the
presence of tachyonic mode on the D-brane world-volume \cite{asone}. As a
result, these systems decay and the decay occurs by a process
known as tachyon condensation \cite{astwo}. Tachyon condensation is well
understood in the open string description using either the string
field theory approach \cite{sz,gs} or the tachyon effective action approach
\cite{kl} on
the brane. However a closed string (or supergravity) understanding
of this process
is far from complete and the purpose of this paper is precisely
to have a closed string understanding of the tachyon condensation.
An earlier attempt in this direction has been made in \cite{bmo} by
giving an interesting interpretation to the previously known \cite{zz,im}
non-supersymmetric, three parameter supergravity solutions with a
symmetry ISO($p,1$) $\times$ SO($9-p$) in ten space-time
dimensions as the coincident D$p$--${\bar {\rm D}}p$ system. The
three parameters in this solution were argued \cite{bmo} to be related
(although the exact relations were not given) to the
physically meaningful parameters, namely, the number of
D$p$-branes ($N$), number of ${\bar {\rm D}}p$-branes (${\bar N}$)
and the tachyon vev\footnote{By tachyon vev we mean the classical value
of the tachyon for which the total energy of the system takes a particular
value which includes the extremum as well as off-shell values \cite{bmo}.}
of the D$p$--${\bar {\rm D}}p$ system.

In this paper we use the static counterpart of the asymptotically
flat time dependent
supergravity solution obtained in \cite{br}.
This is also a non-BPS, three parameter solution with the
symmetry ISO($p,1$) $\times$ SO($9-p$) and is related to the
solution given in ref.\cite{bmo}. This solution can also be naturally
interpreted as the coincident $N$ D$p$-branes and ${\bar N}$
${\bar {\rm D}}p$-branes system given its aforementioned symmetry.
In contrast to the attempt made in \cite{bmo}, we approach the problem
by giving, for the first time, an explicit representation of the
parameters of the solution in terms of $N, \bar N$ and the tachyon
vev of the D$p$--${\bar {\rm D}}p$ system.

We proceed as follows. Once the supergravity solution under
consideration is realized to represent the $N$ D-brane and $\bar
N$ anti D-brane system, we can gain information about the
parameters of the solution by examining how it reduces to a
supersymmetric configuration which either corresponds to a BPS $N$
D$p$-branes (for ${\bar N}=0$) or BPS ${\bar N}$ ${\bar {\rm
D}}p$-branes (for $N=0$) or the final supersymmetric state at the
end of tachyon condensation. We also expect in taking the BPS
limit that only one parameter corresponding to the number of
branes remains and the other parameters of the solution get
automatically removed. For general case when both $N$ and ${\bar
N}$ are non-zero, the solution is not supersymmetric and there
must be  a tachyon on the world-volume of D$p$--${\bar {\rm D}}p$
system belonging to the complex $(N,{\bar N})$ representation of
the gauge group $U(N) \times U({\bar N})$. The end of the tachyon
condensation should give BPS $(N - \bar N)$ D-branes if $N > \bar
N$ or BPS $(\bar N - N)$ anti D-branes if $\bar N > N$. We will
give a general description for arbitrary $N$ and $\bar N$ where $N
= \bar N$ appears as a special case. We will show how the
interplay of the parameters describes the tachyon condensation in
accordance with the conjecture made by Sen \cite{asone,astwo} for
the D$p$--${\bar {\rm D}}p$ system. The recognition of having a
supersymmetric background at the end of the tachyon condensation
is crucial for us to find explicit representation of the
parameters of the solution in terms of $N, \bar N$ and the
tachyon\footnote{When $N={\bar N}= 1$, tachyon is a complex field
and $|T|^2=TT^{\ast}$. But for $N,\,{\bar N}>1$, tachyon is a
matrix and in that case we follow \cite{bmo} to define $|T|^2 =
\frac{1}{N}{\rm Tr}(TT^{\ast})$. Here and in the rest of the paper
we denote $|T|$ as $T$ for simplicity.} vev $T$.

In order to understand the tachyon condensation, we look at the
expression of the total ADM mass of the solution representing the total
energy of the system. We then express this total energy in terms
of the three physical parameters namely, $N$, ${\bar N}$ and $T$ of the
D$p$--${\bar {\rm D}}p$ system using the aforementioned relations. The
total energy can be seen to be equal or less than the sum of the
masses of $N$ D$p$-branes and ${\bar N}$ ${\bar {\rm D}}p$-branes, indicating
the presence of tachyon contributing the negative potential energy to the
system. We will see that the energy expression gives the right picture
of tachyon condensation conjectured by Sen \cite{asone,astwo}. We will
reproduce all the expected results from this general mass formula under
various special limits for $N \neq {\bar N}$ as well as $N = {\bar N}$ at
the top and at the bottom of the tachyon potential. We will also show
how the various known BPS supergravity configurations can be reproduced in
these special limits.

 The static, non-BPS supergravity
$p$-brane solution analogous to the time dependent solution
obtained in \cite{br} has the form in $d=10$,
\bea
ds^2 &=&
F^{-\frac{7-p}{8}} \left(-dt^2 + dx_1^2 + \ldots + dx_p^2\right) +
F^{\frac{p+1}{8}}\left(H\tilde{H}\right)^{\frac{2}{7-p}}
\left(dr^2 + r^2 d\Omega_{8-p}^2\right)\nn e^{2\phi} &=&
F^{-a}\left(\frac{H}{\tilde{H}} \right)^{2\delta}\nn F_{[8-p]} &=&
b\,{\rm Vol}(\Omega_{8-p})
\eea
where we have written the metric
in the Einstein frame. Note that the metric has the required
symmetry and represents the magnetically charged $p$-brane
solution. The corresponding electric solution can be obtained from
(1) by dualizing the field strength $F_{[p+2]} = e^{a\phi} \ast
F_{[8-p]}$, where $\ast$ denotes the Hodge dual. Also in the above
\bea
F &=& \cosh^2\theta \left(\frac{H}{\tilde{H}}\right)^\alpha -
\sinh^2\theta \left(\frac{\tilde{H}}{H}\right)^\beta\nn H &=& 1 +
\frac{\omega^{7-p}}{r^{7-p}}, \qquad \tilde{H}\,\,\,=\,\,\, 1 -
\frac{\omega^{7-p}}{r^{7-p}}
\eea
with the parameter relation
\be
b = (\alpha+\beta)(7 -p) g_s \omega^{7 - p} \sinh2\theta
\ee
Here
$\alpha$, $\beta$, $\theta$, and $\omega$ are integration
constants and $g_s$ is the string coupling. Also $\alpha$ and
$\beta$ can be solved, for the consistency of the equations of
motion,  in terms of $\delta$ as \bea \alpha &=& \sqrt{\frac{2(8 -
p)}{7 - p} - \frac{(7 - p)(p + 1)}{16} \delta^2} +
\frac{a\delta}{2}\nn \beta &=& \sqrt{\frac{2(8 - p)}{7 - p} -
\frac{(7 - p)(p + 1)}{16} \delta^2} - \frac{a\delta}{2}. \eea
These two equations indicate that the parameter $\delta$ is
bounded as \be |\delta| \leq \frac{4}{7 - p} \sqrt{\frac{2 (8 -
p)}{p + 1}}.\ee The solution (1) is therefore characterized by
three parameters $\delta$, $\omega$ and $\theta$. In (1) $a$ is
the dilaton coupling to the $(8-p)$-form field strength and is
given as $a=(p-3)/2$ for the D$p$-branes and $a=(3-p)/2$ for the
NSNS branes. Also $b$ is the magnetic charge and the
Vol($\Omega_{8-p}$) is the volume-form of the $(8-p)$-dimensional
unit sphere. Without any loss of generality, we take $(\alpha +
\beta), b, \theta \geq 0$ as we did already in the above. We note
from (2) that the solution has a curvature singularity at
$r=\omega$ and the physically relevant region is $r>\omega$. As $r
\to \infty$, $H,\, {\tilde H} \to 1$ and so, $F \to 1$. The
solution is therefore asymptotically flat. In order to obtain the
electrically charged solution we dualize $F_{[8-p]}$ in (1) and
get the gauge field $A_{[p+1]}$ as, \be A_{[p+1]} = \sinh\theta
\cosh\theta \left(\frac{C}{F}\right)dx^0 \wedge \ldots \wedge dx^p
\ee where $C$ is defined as, \be C = \left(\frac{H}{{\tilde
H}}\right)^\alpha - \left(\frac{{\tilde H}}{H} \right)^\beta \ee

From the form of the metric in (1), it is clear that the solution
is non-supersymmetric \cite{dkl} because of the presence of
$(H{\tilde H})^{2/(7-p)}$ factor in the last term.  This is also
consistent with our interpretation of the solution as  $N$
D$p$-brane and $\bar N$  ${\bar {\rm D}}p$-brane system. With this
interpretation, we can express the parameter $b$ in terms of $N,
\bar N$ as \be Q_0^p |N-{\bar N}| = \frac{b \Omega_{8-p}}{\sqrt{2}
\kappa_0} \quad \Rightarrow \quad b = \frac{\sqrt{2}\kappa_0 Q_0^p
|N-{\bar N}|} {\Omega_{8-p}} \ee where $Q_0^p =
(2\pi)^{(7-2p)/2}\alpha'^{(3-p)/2}$ is the unit charge on a
D$p$-brane, $\sqrt{2}\kappa_0 = (2\pi)^{7/2}\alpha'^2$ is related
to 10 dimensional Newton's constant and $\Omega_n = 2
\pi^{(n+1)/2}/ \Gamma((n+1)/2)$. Note that $b \to 0$ as $N \to
\bar N$.

For the solution (1), the supersymmetry will be restored if and
only if $H{\tilde H} \to 1$ which always requires $\omega^{7 -p}
\to 0$. We have the following cases for which supersymmetry can be
restored: (1) either $N = 0$ or $\bar N = 0$ or both (the trivial
case); (2) when both $N$ and ${\bar N}$ are non-vanishing susy can
be restored at the end of tachyon condensation. For the second
case, we have two subcases. The first is the familiar one with $N
= \bar N$ for which the end of tachyon condensation should give an
empty spacetime with maximal supersymmetry and the second with $N
\neq \bar N$ for which we expect the final configuration to
represent BPS $(N - \bar N)$ D$p$-branes if $N > \bar N$ or BPS
$(\bar N - N)$ ${\bar {\rm D}}p$-branes if $\bar N > N$. For all
these cases, we expect $\omega^{7 - p} \to 0$. This observation is
crucial to express the three parameters of the solution in terms
of $N, \bar N$ and the tachyon vev $T$. Given the case (1) above
along with the fact that $\omega^{7 -p}$ should be symmetric with
respect to $N$ and $\bar N$, we expect that the leading behavior
of $\omega^{7 - p}$ should be $\omega^{7 - p} \sim (N \bar
N)^\gamma$ with a positive parameter $\gamma$. Keeping other
factors in mind (for example, $\omega^{7 - p} \neq 0$ as $N = \bar
N$),  we make our educated guess for  $\omega^{7 - p}$ as
$\omega^{7 - p} = f (N \bar N)^{1/2}$ with $f$ depending only  on
the tachyon vev $T$ and some other known constants. Considering
case (2) above, we choose $f$ to depend on $T$ as $f \sim \cos T$
with $T = \pi/2$\,\,\,\footnote{At the end of the condensation all
of the eigenvalues of the matrix $T T^\ast$ were argued to be the
same\cite{witten,bmo} as $T_0^2$ and we take $T_0 = \pi/2$ here.}
as the end point of the tachyon condensation. Putting everything
together, $\omega^{7 - p}$ therefore takes the form,
 \be (7 - p) \omega^{7 - p} = \sqrt{\frac{7 -
p}{2 (8 - p)}} (N\bar N)^{1/2} \frac{2 \kappa_0^2}{\Omega_{8 - p}}
T_p \cos T .\ee

   Now we come to determine the parameter $\delta$ which is
expected to be related to $N, \bar N$ and the tachyon vev $T$ as
well. As indicated in eq.(5), the parameter $\delta$ is bounded,
therefore it cannot depend on $N + \bar N$, $N - \bar N$ and/or
$N\bar N$ or any of the inverse powers of them in a simple fashion
since either $N$ or $\bar N$ or both can take arbitrary large or
zero value which cannot give a bounded contribution. Therefore, if
$\delta$ depends on $N, \bar N$ at all, they must appear in such
way that when the terms involving $N$ and $\bar N$ get large they
must cancel each other to give a bounded contribution to $\delta$.
Also when the tachyon vev $T$ takes specific values, the terms
involving $N$ and $\bar N$ should give bounded contribution. Given
the above and considering the bound (5) and the special case of $p
= 3$, our educated guess for $\delta$ is \be \delta =
\frac{a}{|a|} \sqrt{\frac{8 - p}{2(7 - p)}}\left[|a| \sqrt{\cos^2T
+ \frac{(N - \bar N)^2 }{4 N \bar N \cos^2 T}} - \sqrt{a^2
\left(\cos^2 T + \frac{(N - \bar N)^2}{4 N\bar N \cos^2 T}\right)
+ 4 \sin^2 T}\right].\ee With this we can read off $\alpha+\beta$
and $\alpha-\beta$ from (see eq.(4)) \be \alpha+\beta =
2\sqrt{\frac{2(8-p)}{7-p} - \frac{(7-p)(p+1)}{16}\delta^2}, \ee
and \be \alpha-\beta = a \delta, \ee in terms of $N, \bar N$ and
the tachyon vev $T$ explicitly.

With (9) and (11), we can now determine the parameter $\theta$
from (3) as \bea \sinh 2\theta &=& \frac{|N - \bar N|}{c (\alpha +
\beta) (N\bar N)^{1/2} \cos\, T}\nn \cosh 2\theta &=& \sqrt{1 +
\frac{(N - \bar N)^2}{c^2 (\alpha + \beta)^2 N\bar N \cos^2\,
T}},\eea where we have used (8) for $b$ and the constant $c =
\sqrt{(7 - p)/2(8 - p)}$.

So, we have postulated in the above how the two parameters
$\omega$ and $\delta$ are related to $N, \bar N$ and the tachyon
vev $T$ based on the expected properties of the solution and the
characteristic behavior of the tachyon condensation. For given $N$
and $\bar N$, each value of $T$ labels a static solution,
therefore we have a one-parameter family of solutions with $0 \le
T \le \pi/2$. At this point, we cannot be certain that our
educated guess for them is really correct and we have to do some
consistency check. In the following, we will write down the ADM
mass for the solution (1) representing the total energy of the
system, and check whether with the above parameter relations we
can produce all the required properties of the total energy, the
solution as well as the tachyon condensation according to Sen
conjecture.

The total ADM mass of the system can be calculated using the
formula given in \cite{jxl} and for the metric in (1) we obtain, \be
M = \frac{\Omega_{8 - p}}{2 \kappa^2_0} (7 - p) \omega^{7 - p}
\left[(\alpha+\beta) \cosh2\theta + (\alpha-\beta)\right].
 \ee
with $\alpha$ and $\beta$ given in terms of $\delta$ as in eq.
(4). Using (9), (10), and (13) for the parameter $\omega, \delta$
and $\theta$ in terms of $N, \bar N$ and $T$, the mass can be
expressed in a surprisingly simple form as,
\bea M &=& T_p \left(N\bar
N\right)^{1/2} \sqrt{4 \cos^4\, T + \frac{(N - \bar N)^2}{ N\bar N
}}\nn
&=& T_p \sqrt{(N+{\bar N})^2 - 4 N {\bar N}(1-\cos^4 T)}\,\,\,
\leq\,\,\, T_p (N+{\bar N}).
\eea
Thus we note that the total mass is less or equal to the sum of the masses
of $N$ D$p$-branes and ${\bar N}$ ${\bar {\rm D}}p$-branes. The difference
is the tachyon potential energy $V(T)$ which is negative.
One can easily see that $T = 0$
gives the maximum of the energy, therefore the maximum of the
tachyon potential (actually $V(T)=0$ here) while $T = \pi/2$ gives
the corresponding
minima.

Let's check one by one if the above mass
formula produces all required properties of the solution and the
tachyon condensation. At $T = 0$, i.e., at the top of the tachyon
potential, $\cos T = 1$ and we have from the above $M = (N +
\bar N) T_p$, producing the expected result. For this case,
$\delta = 0$, $\alpha = \beta =  \sqrt{2(8 - p)/(7 - p)}$ and
$\omega$ remains finite as can be seen from (9), therefore the
corresponding solution breaks all the susy as it should be. At the end
of the condensation, i.e., $T = \pi/2$, $M = |N - \bar N| T_p$,
again producing the expected result. As $T \to \pi/2$, $\omega \to
0$ and the solution becomes BPS $(N - \bar N)$ D$p$-branes if $N >
\bar N$ or $(\bar N - N)$ ${\bar {\rm D}}p$-branes if $\bar N
> N$.

Let us discuss in a bit detail how the solution behaves at the end of the
condensation. Here $\delta = 0$ for all $p$ except for $p = 3$ (for
which $\delta \to \pm \sqrt{2(8 - p)/(7 - p)}$ as can be seen from
(10)). Therefore we again have from eq.(4) $\alpha = \beta = \sqrt{2(8
- p)/(7 - p)}$ except for $p = 3$. On the other hand for $p=3$,
$\alpha = \beta \to
0$. Even though the parameters $\alpha, \beta$ and $\delta$ are
different for $p \neq 3$ and for $p = 3$, we have a uniform limit
for the function $F$ in (2) at the end of the tachyon condensation as  \be
F \to \bar H = 1 + \frac{\bar\omega^{7 - p}}{r^{7 - p}},\ee with
$\bar\omega^{7 - p} = b/g_s(7 - p)$ finite and $H, \tilde H \to
1$. The corresponding configuration, as can be seen from (1) with
(16), is either BPS $(N - \bar N)$ D$p$-branes if $N
> \bar N$ or  BPS $(\bar N - N)$ ${\bar {\rm D}}p$-branes if $\bar N > N$.

The tachyon condensation can also be seen for the special case of
$N = \bar N$.
In this case for $T = 0$, we have $M = 2 N T_p$ and the
corresponding configuration breaks all the susy while at the end of
the condensation, i.e., at $T = \pi/2$, $M = 0$, corresponding
to an empty spacetime preserving all the susy. This can also be
seen from the configuration (1) since now $\bar\omega^{7 - p} = 0$, therefore
$\bar H = 1$.

Finally let us check whether the mass formula (15) produces the
expected result for $N = 0$ or $\bar N = 0$
or both. We can read off from (15) that $M = N T_p$ when $\bar N = 0$
and $M = \bar N T_p$ when $N = 0$ and $M = 0$ when $N = \bar N =
0$ all as expected. It is not difficult to check from (1) that the
corresponding configuration is either  $N$ BPS D$p$-branes if $\bar N
= 0$ or ${\bar N}$ BPS ${\bar {\rm D}}p$-branes if $N = 0$ or an
empty spacetime when
both $N = 0$ and $\bar N = 0$. The tachyon vev decouples
automatically as expected.

In summary, the supergravity solution (1) is naturally interpreted
as coincident $N$ D$p$-branes and $\bar N$ ${\bar {\rm D}}p$-branes system
given its symmetry and the number of parameters characterizing
this solution. Based on the physical properties of the solution
and the characteristic behavior of tachyon condensation, we give,
for the first time, an explicit representation of the three
parameters of the solution in terms of $N, \bar N$   and the
tachyon vev $T$ which produces all the required properties of the
ADM mass and the solution as discussed in the paper. In this
respect, we capture the right picture of the tachyon condensation
using closed string or supergravity description.

\vs{5}

\noindent {\bf Acknowledgements}

\vs{2}

JXL would like to thank Miao Li for discussion and for reading the
manuscript,  the Michigan Center for Theoretical Physics for
hospitality and partial support during the initial stages of this
work and the CAS Interdisciplinary Center of Theoretical Studies
for hospitality and partial support where part of this work was
completed. He also acknowledges support by grants from the Chinese
Academy of Sciences and the grants from the NSF of China with
Grant No:10245001, 90303002.



\begin{thebibliography}{99}

\bibitem{asone} A. Sen, ``Non-BPS states and branes in string theory'',
hep-th/9904207.

\bibitem{astwo} A. Sen, ``Tachyon condensation on the brane-antibrane
system'',
JHEP 08 (1998) 012, [hep-th/9805170].

\bibitem{sz} A. Sen and B. Zwiebach, ``Tachyon condensation in string
field theory'', JHEP 03 (2000) 002, [hep-th/9912249]; N. Berkovits, A. Sen
and B. Zwiebach, ``Tachyon condensation in superstring field theory'',
Nucl. Phys. B587 (2000) 147, [hep-th/0002211].

\bibitem{gs} A. Gerasimov, S. Shatashvili, ``On exact tachyon potential
in open string field theory'', JHEP 10 (2000) 034, [hep-th/0009103];
D. Kutasov, M. Marino and G. Moore, ``Remarks on tachyon condensation
in superstring field theory'', hep-th/0010108; ``Some exact results
on tachyon condensation in string field theory'', JHEP 10 (2000) 045,
[hep-th/0009148].

\bibitem{kl} P. Kraus and F. Larsen, ``Boundary string field theory of
the DD-bar system'', Phys. Rev. D63 (2001) 106004, [hep-th/0012198];
T. Takayanagi, S. Terashima and T. Uesugi, ``Brane-antibrane action
from boundary string field theory'', JHEP 03 (2001) 019, [hep-th/0012210].

\bibitem{bmo} P. Brax, G. Mandal and Y. Oz, ``Supergravity description of
non-BPS branes'', Phys. Rev. D63 (2001) 064008, [hep-th/0005242].

\bibitem{zz} B. Zhou and C. Zhu, ``The complete black-brane solutions in
D-dimensional coupled gravity system'', hep-th/9905146.

\bibitem{im} V. Ivashchuk and
V. Melnikov, ``Exact solutions of multidimensional gravity with antisymmetric
forms'', Class. Quant. Grav. 18 (2001) R87, [hep-th/0110274].

\bibitem{br} S. Bhattacharya and S. Roy, ``Time dependent supergravity
solutions in arbitrary dimensions'', JHEP 12 (2003) 015, [hep-th/0309202].

\bibitem{dkl} M. J. Duff, R. R. Khuri and J. X. Lu, ``String solitons'',
Phys. Rept. 259 (1995) 213, [hep-th/9412184].

\bibitem{witten} E. Witten, ``D-branes and K-theory", JHEP 12
(1998) 01, [hep-th/9810188].

\bibitem{jxl} J. X. Lu, ``ADM masses for black strings and p-branes'', Phys.
Lett. B313 (1993) 29, [hep-th/9304159].

\end{thebibliography}
\end{document}